\documentclass[two column,prb,superscriptaddress]{revtex4}
\usepackage{bm}
\usepackage{graphics}
\usepackage{graphicx}
\usepackage{amsfonts}
\usepackage{amsmath}
\usepackage{amssymb}
\usepackage{graphicx}
\usepackage{color}

\begin{document}

\preprint{HEP/123-qed}
\title[Short title for running header]{High Quantum Efficiency Ultrananocrystalline Diamond
Photocathode:\\
Negative Electron Affinity Meets $n$-doping}

\author{K.J. P\'{e}rez Quintero}
\affiliation{Nanoscience and Technology Division, Argonne National
Laboratory, Argonne, IL 60439, USA} \affiliation{Physics
Department, University of Puerto Rico,San Juan, PR 00931, USA}

\author{S. Antipov}
\affiliation{Euclid TechLabs, Solon, OH 44139, USA}
\affiliation{High Energy Physics Division, Argonne National
Laboratory, Argonne, IL 60439, USA}

\author{A.V. Sumant}
\email{sumant@anl.gov} \affiliation{Nanoscience and Technology
Division, Argonne National Laboratory, Argonne, IL 60439, USA}

\author{C. Jing}
\affiliation{Euclid TechLabs, Solon, OH 44139, USA}
\affiliation{High Energy Physics Division, Argonne National
Laboratory, Argonne, IL 60439, USA}

\author{A.D. Kanareykin}
\affiliation{Euclid TechLabs, Solon, OH 44139, USA}

\author{S.V. Baryshev}
\email{sergey.v.baryshev@gmail.com} \affiliation{Euclid TechLabs,
Solon, OH 44139, USA} \affiliation{High Energy Physics Division,
Argonne National Laboratory, Argonne, IL 60439, USA}

\begin{abstract}
We report results of quantum efficiency (QE) measurements carried
out on a 150 nm thick nitrogen-incorporated ultrananocrystalline
diamond terminated with hydrogen; abbreviated as (N)UNCD:H.
(N)UNCD:H demonstrated a QE of $\sim$10$^{-3}$ ($\sim$0.1\%) at
254 nm. Moreover, (N)UNCD:H was sensitive in visible light with a
QE of $\sim$5$\times$10$^{-8}$ at 405 nm and
$\sim$5$\times$10$^{-9}$ at 436 nm. After growth and prior to QE
measurements, samples were exposed to air for about 2 hours for
transfer and loading. Such design takes advantage of a key
combination: 1) H-termination inducing negative electron affinity
(NEA) on the (N)UNCD and stabilizies its surface against air
exposure; and 2) N-incorporation inducing $n$-type conductivity in
intrinsically insulating UNCD.
\end{abstract}

\maketitle

The photocathode is a key component of the electron injectors in
synchrotrons, free electron lasers, linear accelerators (linacs),
and ultrafast electron systems for imaging and diffraction. Choice
of a specific photocathode is application specific, and there is
always a trade-off: QE vs. lifetime/robustness vs. response time
vs. emittance. It is generally accepted that if a technology
providing a high QE photocathode operating at moderate vacuum
conditions existed, it would greatly benefit the field of
photoinjectors R\&D.\cite{1}

Semiconductor photocathodes still hold records in terms of QE.
These are low work function (WF) alkali/multialkali based
materials that are used in a form of thin films to absorb light
and emit electrons,\cite{2,3} or in a form of ultrathin layers to
activate traditional metal photocathodes.\cite{4} Activation of
heavily doped $p$-Si or $p$-GaAs surfaces with alkali Cs has led
to a special photocathode type with negative electron affinity
(NEA). NEA is a unique circumstance, when electrons injected to
the conduction band can be emitted directly into the vacuum. Such
NEA photocathodes are bright electron sources because of their
high QE and low emittance, which decreases as the NEA value
increases.\cite{5} The NEA value is a measure of how the low
vacuum level locates with respect to conduction band minimum.
Nevertheless, the main drawback of alkali-based photocathodes
remains the same -- they require a vacuum base pressure
$\leq$10$^{-10}$ Torr for synthesis, handling and operation.

Wide bandgap ($>$5 eV) semiconductors are another class of NEA
materials. This includes AlN, BN, and diamond.\cite{6,7} In
diamond, NEA can be either an inherent surface property\cite{8} or
an engineered one\cite{9} via surface treatment in a hydrogen
environment. Since the first experiment which demonstrated a
remarkable quantum yield from a NEA diamond surface under vacuum
UV illumination,\cite{8} prototypes of solar blind high efficiency
photocathodes for space research detectors have been
introduced.\cite{10} High purity H-terminated synthetic diamond
has been found to be an excellent electron amplifier, where the
primary electrons from a standard QE photocathode (e.g. Cu)
accelerated to a keV energy get multiplied upon transmission
through a thin diamond film. Chang \emph{et al.}\cite{11} have
demonstrated gain coefficients as high as 200. In most previous
applications, high purity (no dopants) diamonds or boron doped
($p$-type conductivity) diamonds were used. Thus vacuum UV
wavelengths ($<$200 nm) were targeted. Boron $p$-doping did not
play a significant role\cite{12} as the boron level is only 0.4 eV
above the top of the valence band in diamond.

To take advantage of NEA and sensitize diamond towards the near UV
and visible spectral ranges, and thus make it of interest to the
photoinjectors R\&D community, one should introduce electrons in
diamond. A way to do so would be $n$-doping. Relatively recent
progress in $n$-doping of micro-, nano- and ultranano-crystalline
diamond offers a few options: sulfur (activation energy,
$\varepsilon_a$, 0.4 eV\cite{13}), phosphorous
($\varepsilon_a$=0.6 eV\cite{14}), and nitrogen
($\varepsilon_a$=1.7 eV\cite{15}). Given that the electron
affinity promoted by hydrogen can be as low as --1 eV (NEA value=1
eV),\cite{16} all aforementioned dopants are capable of promoting
visible light photo-excitation. To date, there is one report
experimentally showing that (N)UNCD:H is sensitive to the visible
range. Sun \emph{et al.}\cite{17} reported a measurable external
quantum effect at room temperature between 400 and 480 nm; but no
QE values were presented. With this letter, we report
proof-of-concept QE measurements suggesting that $n$-doped UNCD:H
is an emergent air resistant NEA photocathode for photoinjectors.
QE measurements were carried out in the near UV range 250-270 nm,
standard for many photocathode applications, and in visible light
at 405 and 436 nm. The cathode was exposed to air for about 2
hours for transfer and loading; QE was measured at base pressure
$\sim$10$^{-6}$ Torr.

(N)UNCD films were synthesized on a polycrystalline molybdenum
substrate in a 915 MHz microwave-assisted plasma chemical vapor
deposition (MPCVD) reactor (Lambda Technologies Inc.) Growth of
UNCD on foreign substrates requires a nanodiamond (ND) pre-seeding
treatment prior to deposition to promote growth in the MPCVD.
Slurry of ND particles from Ad\'{a}mas Technologies was used. The
average particle size of the seeds was 5-10 nm. Mo substrates were
immersed into the ND slurry and subject to ultrasonic treatment in
the solution for 20 minutes. Subsequent growth of the (N)UNCD
films was under following conditions: substrate temperature 850
$^\circ$C; operation chamber pressure 56 Torr; microwave power 2.3
kW; individual gas flows in the precursor gas mixture were 3 sccm
CH$_4$/160 sccm Ar/40 sccm N$_2$. Fig.1a shows a scanning electron
micrograph of a deposited film taken by an FEI Nova 600 NanoLab. A
uniform microstructure, typical for UNCD, was observed. Fig.1b
represents a Raman spectrum recorded by a Renishaw InVia Raman
Microscope using a 633 nm laser line. The shoulder around 1140
cm$^{-1}$ corresponds to the $\nu_1$ (C-H in-plain bending)
vibrational mode of trans-polyacetylene and the shoulders at 1340
and 1540 cm$^{-1}$ correspond to the D and G bands of diamond,
respectively.\cite{18,19} The films had a conductivity comparabale
with that of Mo substrates. As a final step, the samples underwent
to H-termination procedure for 15 minutes. It was accomplished in
the same MPCVD reactor at substrate temperature of 750 $^\circ$C.
H$_2$ gas flow was 200 sccm at chamber pressure 15 Torr, and the
microwave power was 2 kW. After that, the samples were left to
cool down to room temperature naturally.

\begin{figure}[t] \centering
\includegraphics[width=5cm]{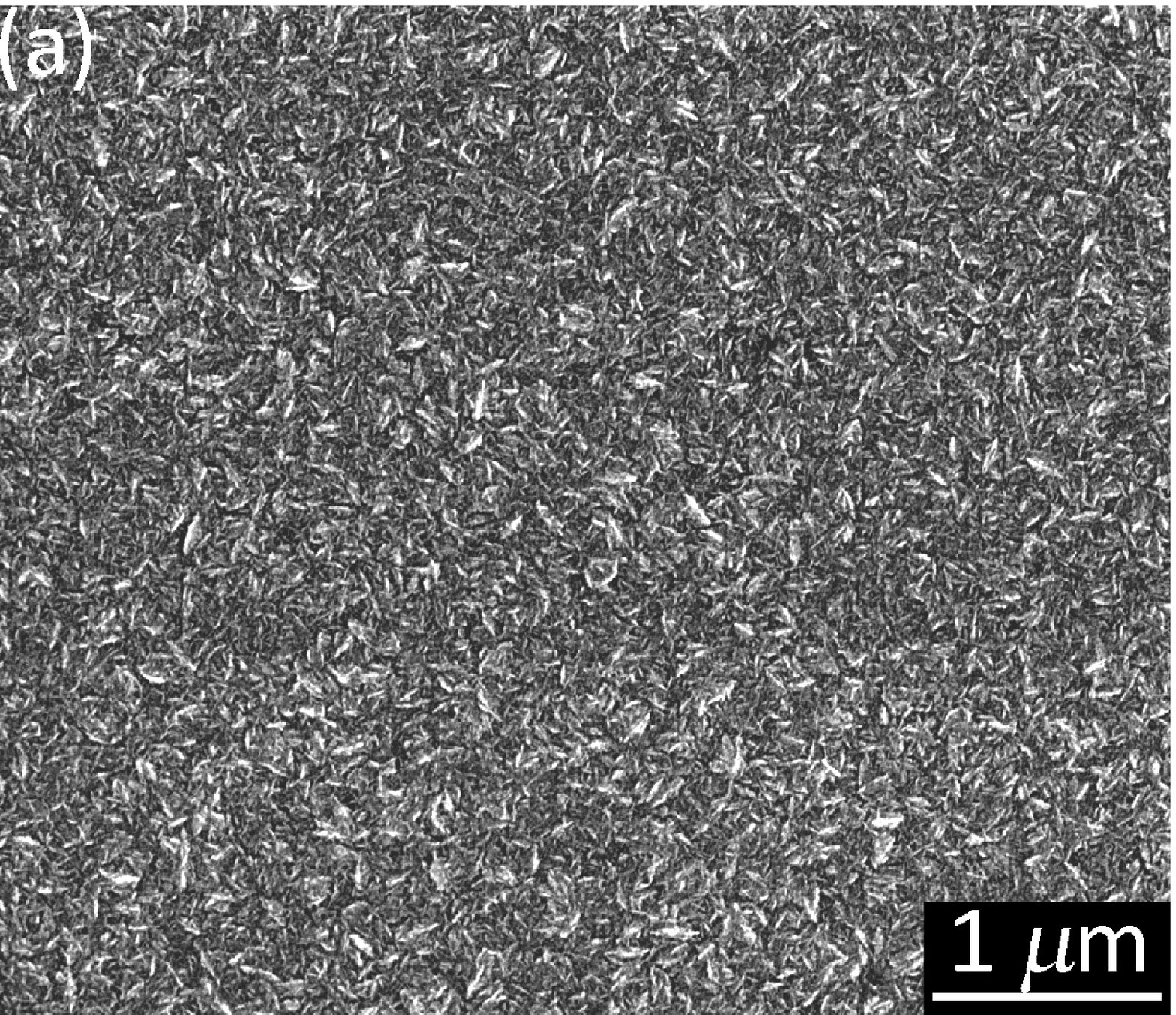}
\includegraphics[width=6.1cm]{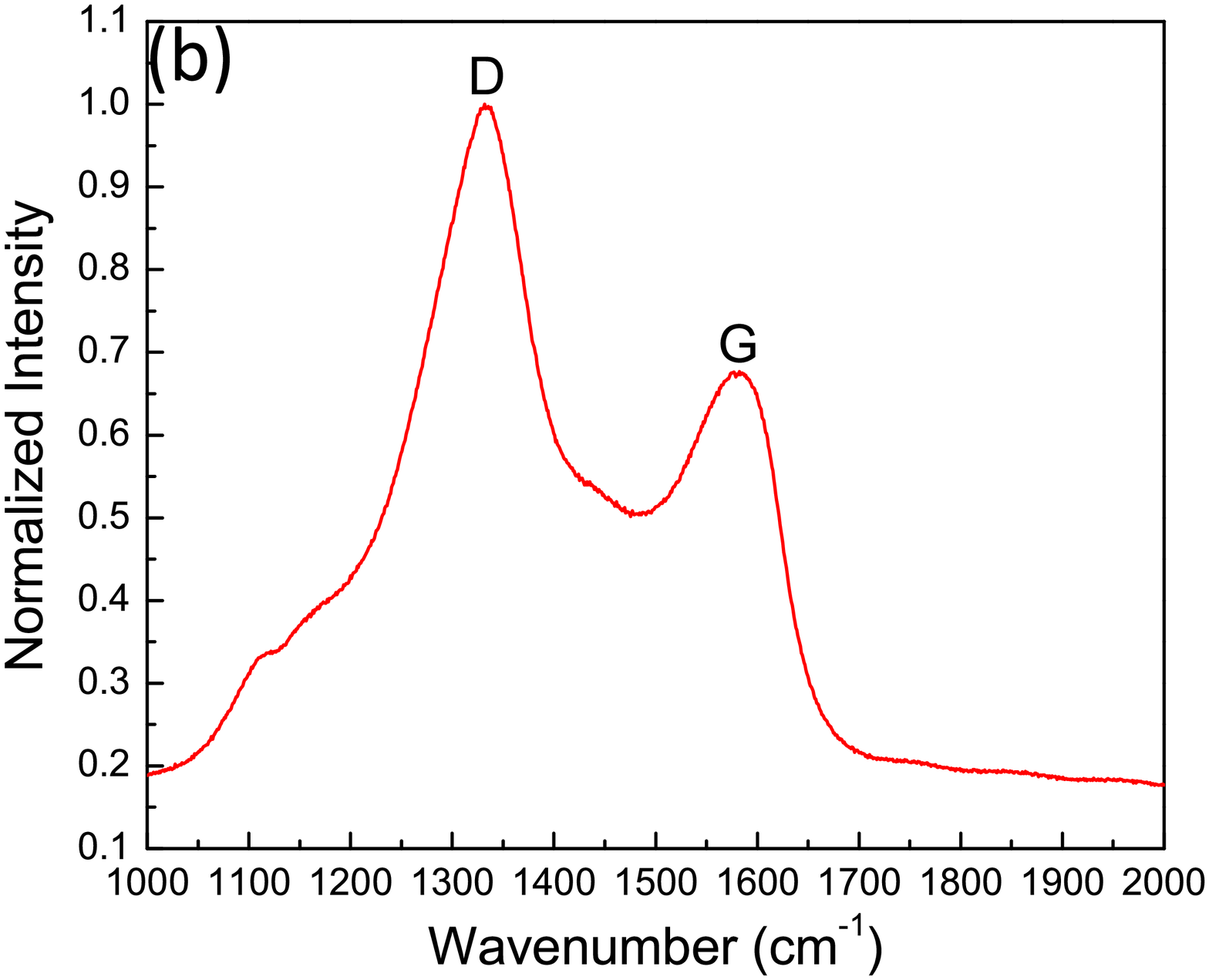}
\caption{(a) SEM surface topography and (b) Raman spectrum typical
for (N)UNCD films on molybdenum.}
\end{figure}

Measurements of the synthesized samples were performed in a
commercial Kelvin probe (KP) instrument (KP6500 from McAllister
Technical Service) with custom in-house modifications so that the
WF and QE can be obtained in the same experimental run. Before or
after termination, all samples were taken from the synthesis
chamber and transported to the KP under ambient conditions; total
exposure time was about 2 hours. The KP chamber in all
measurements was evacuated to a base pressure of $\sim$10$^{-6}$
Torr. Fig.2a represents a cartoon of the experimental setup. A
voltage of +300 V was applied to a small aluminum anode plate, and
a current of photoelectrons to the ground was collected by the
same source/ammeter (Keithley 6487) with a threshold sensitivity
of $\pm$10 fA. The anode plate was introduced into a KP chamber at
an angle such that it is does not interfere with light beam and a
tip assessing WF. A sample holder actuator and the KP tip are both
retractable, and ideal positions can be found for QE and WF
measurements independently. WFs for (N)UNCD samples were
determined by KP with respect to its calibrated tip (WF=4.6 eV)
before and after they underwent H-plasma treatment. A sample
holder made of standard polycrystalline copper was used as a
reference. All deduced WF values are plotted in Fig.2b. WF
dependence on time is a standard representation for KP to estimate
the signal's noise and drift to get a confident WF measurement.

\begin{figure}[t] \centering
\includegraphics[width=4cm]{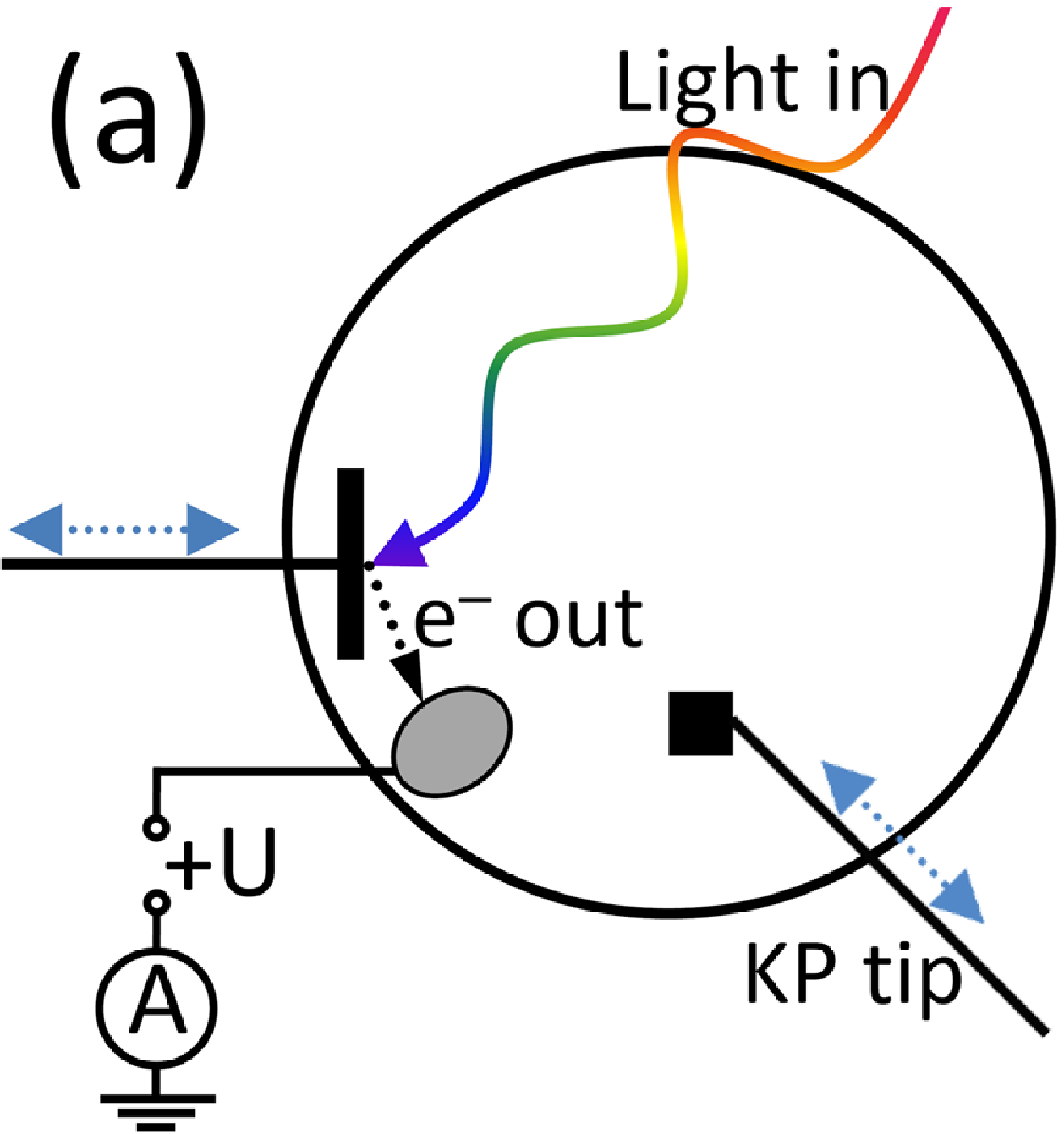}
\includegraphics[width=6.1cm]{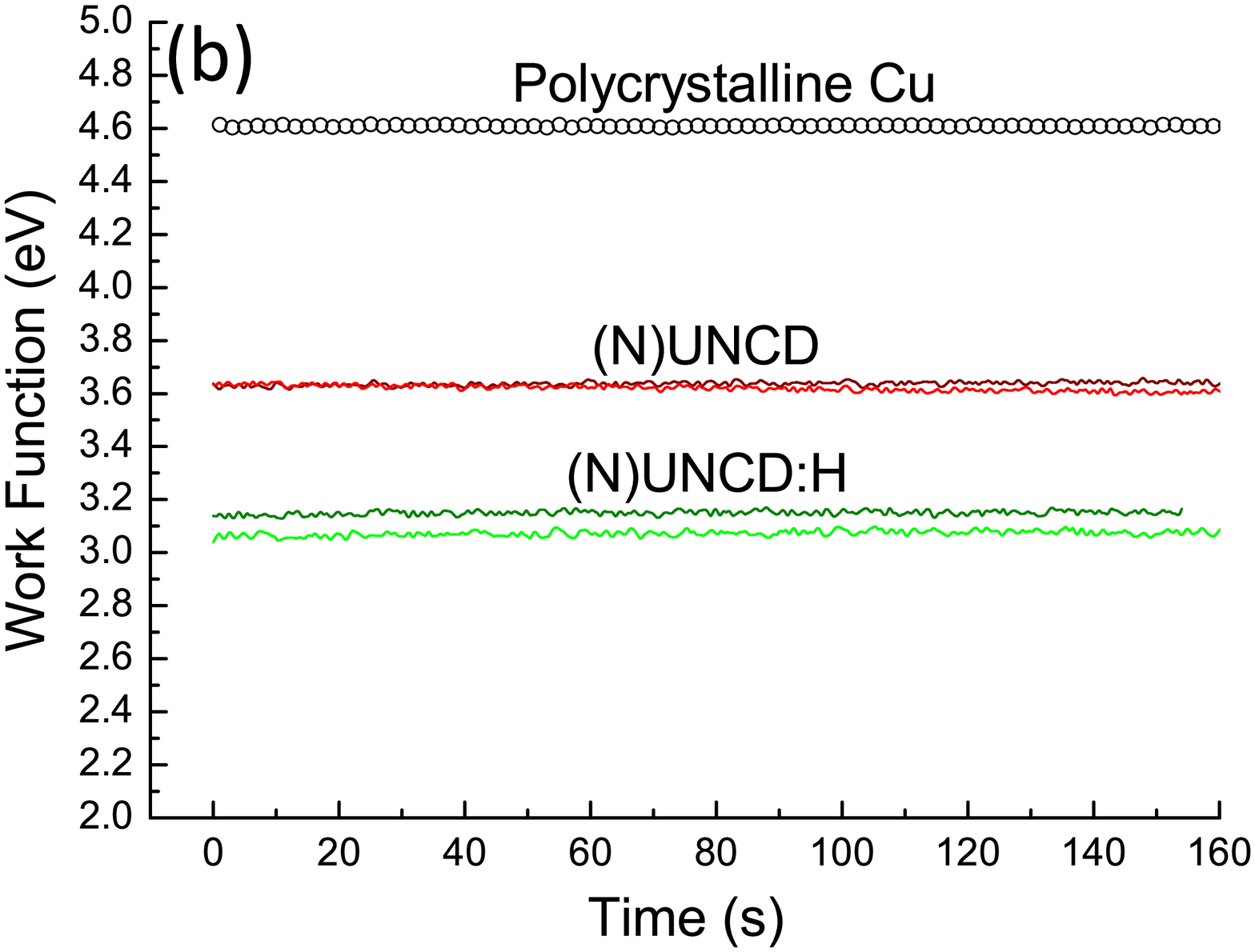}
\caption{(a) A schematic top view of the modified KP chamber; (b)
measured WFs of (N)UNCD before and after H-termination (two
measurements for each case), and a copper WF as a reference.}
\end{figure}

QE measurements were performed using an arc broadband Hg lamp
(Spectra-Physics/Newport Oriel Instruments series 66900) as a
light source. A light spot size from the source was adjusted by an
aperture and focused by a lens; spot size on sample's surface was
$\sim$1 mm$^2$. A number of Newport filters were used to define a
spectral dependence of (N)UNCD QE before and after H-termination,
namely, 254, 313, 365, 405, and 436 nm. The output power of the
lamp at each filtered wavelength was assessed by a calibrated
power meter (Ophir Nova II), equipped with a calibrated photodiode
(Ophir PD300-UV). The photoelectron current was recorded at each
wavelength, and QEs were calculated. All numbers are compiled and
plotted in Fig.3.

As expected, upon $n$-doping and H-termination, UNCD sensitivity
shifted toward near UV/visible wavelengths. There are two main
features in Fig.3 we would like to stress. The first feature is QE
in the band 250-270 nm which is of common interest to the
photocathode community. QE of the originally grown (N)UNCD was
5.3$\times$10$^{-6}$. Given the measured WF 3.6 eV, it is a quite
moderate effect compared to the single crystal Cu (100) QE of
5$\times$10$^{-5}$ with WF=4.2 eV.\cite{21} Remarkably, the QE was
enhanced by a factor of 140 upon H-termination, setting (N)UNCD at
the low boundary of a QE range of alkali-based photocathodes.
Secondly, diamond films were sensitive in visible blue. KP results
suggest that in all cases photoemission was in the \emph{sub-WF}
regime. For (N)UNCD at 365 and 405 nm and for (N)UNCD:H at 436 nm,
this seems a plausible conclusion. It can be explained by emission
from grain boundaries with a lowered WF, caused by the local
environment,\cite{22} accounted also for strong field emission
from flat polycrystalline diamond surfaces.\cite{23} Photoemission
from (N)UNCD:H in visible blue at 405 nm is most probably a
regular threshold process -- photon energy of 3.06 eV versus WF
3.07$\pm$0.01 eV and 3.15$\pm$0.01 eV as determined by KP (light
green and olive solid lines in Fig.2b). In any of the 2 regimes,
incorporation of nitrogen lead to sustainable currents from UNCD
surfaces in the blue range.

\begin{figure}[t] \centering
\includegraphics[width=7.7cm]{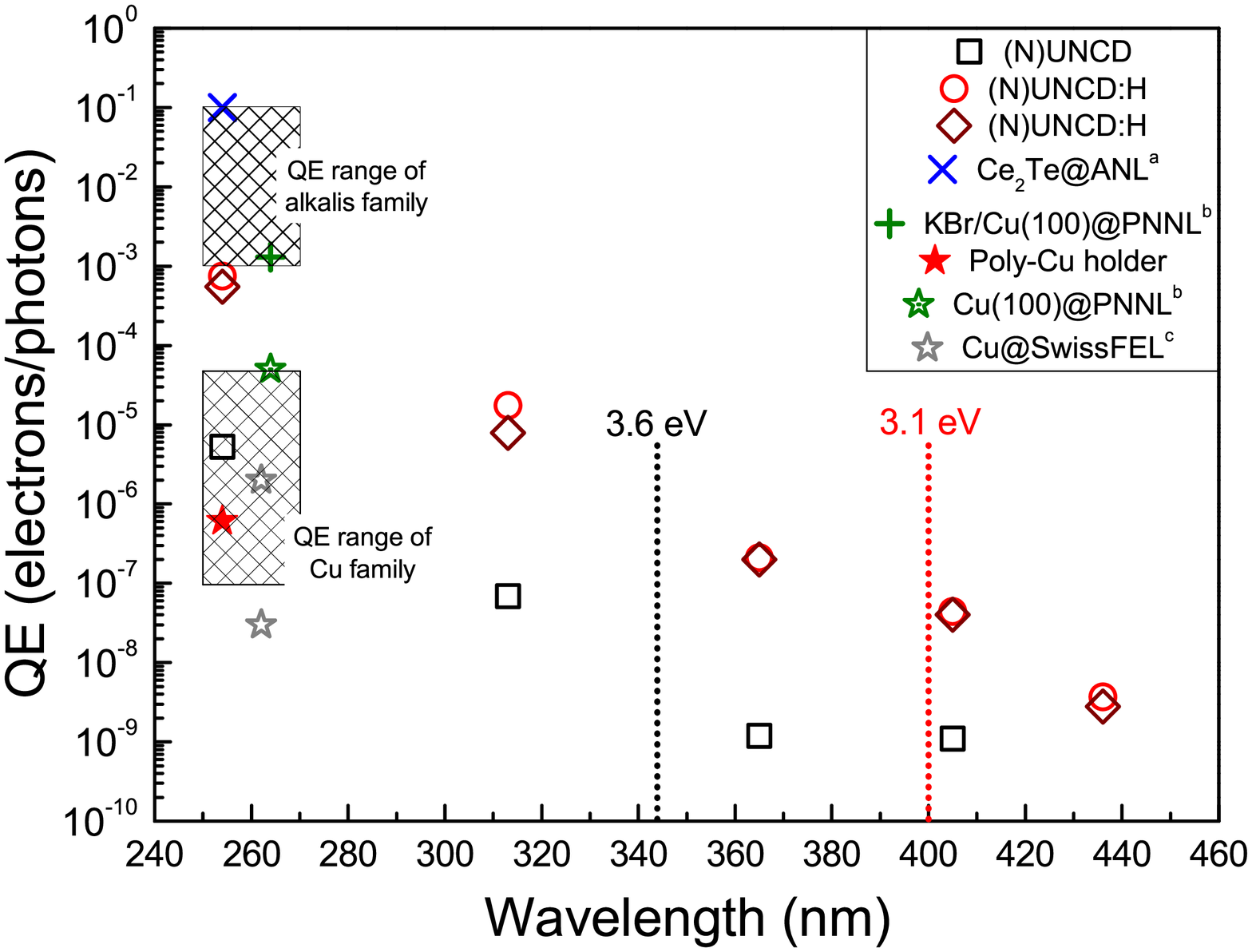}
\caption{Summary of the experimental QEs: one measurement before
and two measurements after termination. Some reference data are
plotted to clearly emphasize the QE effects in the (N)UNCD:H
system. Black and red dotted lines are WFs determined for (N)UNCD
and (N)UNCD:H by KP respectively. A QE of a Cu sample holder (its
WF is in Fig.2b) used as a reference (red solid star).$^a$Ref.[2];
$^b$Ref.[21]; $^c$Ref.[24].}
\end{figure}

In conclusion, by combining hydrogen passivation of the surface,
lowering the work function, and nitrogen introducing electron
states in the band gap close to the conduction band minimum, a
proof-of-concept was demonstrated that ultrananocrystalline
diamond is an emergent robust high efficiency photocathode. This
was accomplished by measuring a QE dependence on wavelength of
primary photons. (N)UNCD:H films of 150 nm thickness had a QE of
$\sim$10$^{-3}$ at 254 nm, and were sensitive in the visible range
(between 405 and 436 nm). It is reasonable to expect that QE in
near UV and sensitivity in the visible, toward 532 nm, can be
further increased. A route to achieve this requires detailed
investigation and optimization of: 1) thickness for the best
photon absorption; 2) defect engineering in the band gap to find
the best trade-off between donors' activation energy and donors'
concentration effecting simultaneously the density of states and
electron lifetime; and 3) defect engineering on the surface to
avoid any upward band bending and to achieve work functions
compared with $n$-dopant's activation energies.

We thank Robert Nemanich and Franz Koeck (ASU) for valuable
discussions, and Eric Wisniewski and Zikri Yusof (IIT) for
technical assistance. Euclid TechLabs acknowledges support from
DOE SBIR, grant DE/SC0009572. Samples were synthesized at the CNM
ANL, a U.S. DOE, Office of Science, Office of Basic Sciences User
Facility under contract DE-AC02-06CH11357. Funding was provided,
in part, by NASA EPSCoR grant NNX13AB22A and NASA Space Grant
NNX10AM80H.


\begin{thebibliography}{}

\bibitem{1} D.H. Dowell \emph{et al.}, NIM A \textbf{622}, 685 (2010).

\bibitem{2} E.E. Wisniewski \emph{et al.}, NIM A \textbf{711}, 60 (2013).

\bibitem{3} T. Vecchione \emph{et al.}, Appl. Phys. Lett. \textbf{99}, 034103
(2011).

\bibitem{4} J. Maldonado \emph{et al.}, Phys. Rev. STAB \textbf{11}, 060702
(2008).

\bibitem{5} S. Karkare \emph{et al.}, Phys. Rev. Lett. \textbf{112}, 097601
(2014).

\bibitem{6} M.J. Powers \emph{et al.}, Appl. Phys. Lett.
\textbf{67}, 3912 (1995).

\bibitem{7} R.J. Nemanich \emph{et al.}, Diam. Relat. Mater. \textbf{5}, 790
(1996).

\bibitem{8} F. Himpsel \emph{et al.}, Phys. Rev.
B \textbf{20}, 624 (1979).

\bibitem{9} J. van der Weide \emph{et al.}, Phys. Rev. B \textbf{50}, 5803 (1994).

\bibitem{10} A.S. Tremsin \emph{et al.}, SPIE Proc. \textbf{4139}, 16 (2000).

\bibitem{11} X. Chang \emph{et al.}, Phys.
Rev. Lett. \textbf{105}, 164801 (2010).

\bibitem{12} A. Laikhtman \emph{et al.}, Appl. Phys. Lett.
\textbf{73}, 1433 (1998).

\bibitem{13} P. Kulkarni \emph{et al.}, J. Appl. Phys. \textbf{103}, 084905 (2008).

\bibitem{14} M. Nesladek, Semicond. Sci. Technol. \textbf{20}, R19 (2005).

\bibitem{15} H.B. Dyer \emph{et al.}, J. Chem. Phys. \textbf{42}, 1898 (1965).

\bibitem{16} J. Cui \emph{et al.}, Phys. Rev. Lett. \textbf{81}, 429
(1998).

\bibitem{17} T. Sun \emph{et al.}, Appl. Phys.
Lett. \textbf{99}, 202101 (2011).

\bibitem{18} I.I. Vlasov \emph{et al.}, Phys. Stat. Sol. A \textbf{203}, 3028 (2006).

\bibitem{19} H. Kuzmany \emph{et al.}, Carbon
\textbf{42}, 911 (2004).

\bibitem{20} J.S. Kim \emph{et al.}, Synthetic Metals
\textbf{111-112}, 311 (2000).

\bibitem{21} W. He \emph{et al.}, Appl. Phys. Lett.
\textbf{102}, 071604 (2013).

\bibitem{22} V. Chatterjee \emph{et al.}, Appl.
Phys. Lett. \textbf{104}, 171907 (2014).

\bibitem{23} K. Okano \emph{et al.}, Nature \textbf{381}, 140 (1996).

\bibitem{24} F. Le Pimpec \emph{et al.}, J. Vac. Sci. Technol. A \textbf{28}, 1191 (2010).

\end{thebibliography}
\end{document}